\documentclass[preprint2]{aastex}
\usepackage{longtable}
\usepackage{rotating}
       
\begin{document}
 
 \title{\bf{A proper motions study of the globular cluster NGC~3201.\thanks{Based on
   observations with the MPG/ESO  2.2m and ESO/VLT telescopes, located
   at  La Silla  and  Paranal Observatory,  Chile,  under DDT  programs
   164.O-0561(F), 093.A-9028(A) and the archive material.}}}
 
 \author{Devesh P. Sariya$^1$, Ing-Guey Jiang$^1$, R. K. S. Yadav$^2$} 
 
 \affil{
 {$^1$Department of Physics and Institute of Astronomy,}\\ 
 {National Tsing-Hua University, Hsin-Chu, Taiwan}\\
 {email: deveshpath@gmail.com}\\
 {$^2$Aryabhatta Research Institute of Observational Sciences,}\\
{Manora Peak, Nainital 263 002, India}\\
}
 
\begin{abstract}
      With a high value of heliocentric radial velocity, a retrograde orbit, and being
      suspected to have an extragalactic origin, NGC~3201 is an interesting
      globular cluster for kinematical studies. 
      Our purpose is to calculate the relative proper
      motions (PMs) and membership probability for the stars in the wide region of
      globular cluster NGC 3201. Proper motion based membership probabilities
      are used to isolate the cluster sample from the field stars.
      The membership catalogue will help address the question of chemical
      inhomogeneity in the cluster.
      Archive CCD data taken with a wide-field imager (WFI) mounted on the ESO {\bf 2.2}m
      telescope are reduced using the high-precision astrometric software
      developed by Anderson et al. for the WFI images. The epoch gap between the two
      observational runs is $\sim$14.3 years. To standardize the $BVI$ photometry,
      Stetson's secondary standard stars are used.
      The CCD data with an epoch gap of $\sim$14.3 years enables us to decontaminate
      the cluster stars from field stars efficiently. The median precision of PMs is better than
      $\sim$0.8 mas~yr$^{-1}$ for stars having $V<$18 mag that increases up to
      $\sim$1.5 mas~yr$^{-1}$ for stars with $18<V<20$ mag. Kinematic membership
      probabilities are calculated using proper motions for stars brighter than
      $V\sim$20 mag. An electronic catalogue of positions, relative PMs, $BVI$
      magnitudes and membership probabilities in $\sim$19.7$\times$17 arcmin$^2$
      region of NGC 3201 is presented. We use our membership catalogue to
      identify probable cluster members among the known variables and
      $X$-ray sources in the direction of NGC 3201.
\end{abstract}
 
 
\section{Introduction}
\label{Intro}
Galactic globular clusters are very important for the study of the
halo and bulge regions of our Galaxy. In particular, they are the best tools to understand
the kinematics and dynamics of the halo region of the Milky Way by the virtue of
being easily distinguishable at large distances (Cudworth 1997, Dambis 2006).
Since the emergence of CCDs, proper motion (PM) studies can be carried out
with unprecedented precision
using CCD data with smaller epoch differences than data from photographic plates.
Proper motions are the root to learn about the kinematics and orbit of the clusters
as well as providing kinematical membership probabilities of the stars.
Membership status is often pivotal to spectroscopic studies, to avoid
observing field stars lying in the cluster's field (Cudworth 1986).
At fainter magnitudes, in particular, field stars dominate, and proper motions
become very important in removing them from the sample (Piotto et al. 2004).

NGC 3201 is a sparse, metal-poor, intermediate-mass halo globular cluster.
The basic cluster parameters of NGC 3201 taken from
Harris (1996, 2010 edition) are listed in Table~\ref{par}.
Since the cluster is less centrally concentrated
than most other globular clusters, even ground based telescopes can
probe its central region. Due to this advantage and its proximity, NGC 3201 has
been studied extensively for photometric studies
(Menzies 1967, Alcaino 1976,
Lee 1977, Alcaino et al. 1981,1989, Cacciari 1984a, 1984b, Penny 1984,
Brewer et al. 1993, Covino \& Ortolani 1997, Kravtsov et al. 2009 etc.).
von Braun \& Mateo (2001) presented an extinction map for NGC 3201 and explained that
due to its low Galactic latitude, the effect of differential reddening across this globular
cluster is very significant.
NGC 3201 has been studied for the question of chemical abundances and possibility of
inhomogeneity in its stellar population by several authors
(e.g., Chun 1988, Gonzalez \& Wallerstein 1998, Covey et al. 2003, Kravtsov et al. 2010,
Simmerer et al. 2013, Mu{\~n}oz et al. 2013,  Mucciarelli et al. 2015 and references therein).
NGC 3201 is known to harbor many variable stars including RR Lyrae, SX Phoenicis etc.
It belongs to Oosterhoff type I according to Oosterhoff dichotomy of RR Lyrae Stars.
The cluster has been the subject of many investigations for searching and characterizing its variable stars
(e.g., Lee \& Carney 1999, von Braun \& Mateo 2002, Piersimoni et al. 2002, Mazur et al. 2003,
Layden et al. 2003, Arellano Ferro et al. 2014, Kaluzny et al. 2016 and references therein).
Webb et al. (2006) studied NGC 3201 using $XMM-Newton$ $X$-ray observatory.

NGC 3201 is a very interesting globular cluster for its kinematical features.
It shows a very high value of heliocentric radial velocity (494.2 km/sec, Cote et al. 1994)
suggesting a retrograde orbit about the Galactic center 
(Casetti-Dinescu et al. 2007). It has been suspected to have an extragalactic origin
and has probably been accreted by the Milky Way.
Dynamics of the cluster was studied by Da Costa et al. (1993) and Cote et al. (1994, 1995).
The Radial Velocity Experiment (RAVE) catalogue was used to study this cluster by
Kunder et al. (2014) and Anguiano et al. (2015, 2016).
Anguiano et al. (2016) presented the distribution of stars based on UCAC4 proper motions.
Chen \& Chen (2010) state that 
NGC 3201 appears to have passed through the Galactic disk a few Myr ago 
and the cluster has clumps along its Galactic north-south axis.

Casetti-Dinescu et al. (2007) provided the absolute proper motion of NGC 3201
($\mu_{\alpha}cos\delta=5.28\pm0.32$ mas yr$^{-1}$, $\mu_{\delta}=-0.98\pm0.33$ mas yr$^{-1}$)
using a combination of photographic plate data with CCD.
Recently, Zloczewski et al. (2012) (hereafter, Zl12) have determined proper motions for
the stars in the region of NGC 3201 and provided membership probabilities in the central region of the cluster.

From the above discussion, it is noticeable that this kinematically
interesting globular cluster is not well studied for proper motions over a wider region.
As reported by Brewer et al. (1993) and Kravtsov et al. (2009), the studies of NGC 3201 are
obstructed by significant field star contamination. Wide Field Imagers (WFIs)
enable us to cover the broad regions of star clusters, sometimes up to their tidal trails.
The archive data observed with WFI@2.2m  at La Silla, Chile has been used previously to
provide proper motions using a time gap of a few years
(Anderson et al. 2006; Yadav et al. 2008, 2013; Bellini et al. 2009; Sariya et al. 2012, 2015).
NGC 3201, being a sparse cluster and the available epoch gap being $\sim$14.3 years in the archive data of WFI@2.2M 
allows us to determine precise proper motions over a broad region of the cluster.

The main goal of the present article is to provide relative PMs and membership probabilities ($P_\mu$)
for stars having visual magnitudes up to 20 mag in the wide field of NGC 3201.
We also provide an electronic catalogue for 8322 stars which contains
$B, V, I$ magnitudes, PMs and membership probabilities for the follow-up studies of the cluster.
Our membership catalogue covers a region of $\sim$19.7$\times$17 arcmin$^2$
which is wider than the area covered in the PM study by 
Zl12 ($\sim$14.6$\times$9.7 arcmin$^2$).
Our membership catalogue will be helpful to select the more likely cluster
members while addressing the question of chemical inhomogeneity which has been 
an intriguing aspect of this cluster.

Information about the data used and reduction procedures are described in
Section~\ref{OBS} where we also discuss PMs, vector point diagrams (VPDs)
with color-magnitude diagram (CMD) of the cluster.
Cluster membership analysis is provided in Section~\ref{MP}.
We use our membership catalogue to examine the membership status of earlier reported variables
and $X$-ray sources in Section~\ref{app}.
The electronic catalogue being presented for the 
further studies of the cluster
is explained in Section~\ref{catl}.
Conclusions follow in Section~\ref{con}.

\begin{figure}
\centering
\includegraphics[width=8.5cm]{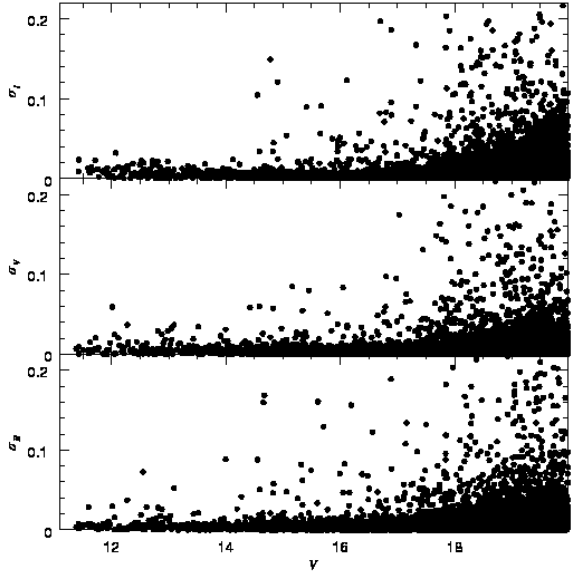}
\caption{Plot of the rms of the residuals around the 
mean $B,V$, and $I$ magnitudes, for stars in the NGC 3201 field imaged
in this study, plotted against V magnitude}
\label{errormag}
\end{figure}
\begin{figure}
\centering
\includegraphics[width=8.5cm]{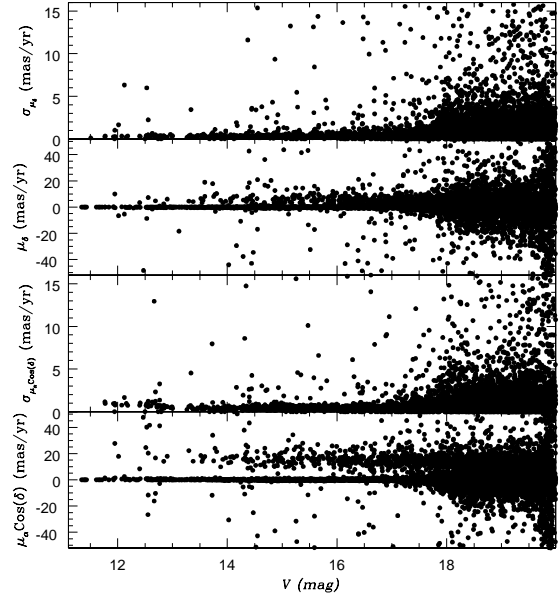}
\caption{Proper motions of stars in the NGC 3201 field, in the globular
cluster reference frame. Proper motions in RA and Dec are plotted
against $V$ magnitude, along with their rms errors}.
\label{errorpm}
\end{figure}
\begin{table}
\caption{Basic parameters of NGC~3201 taken from Harris (1996, 2010 edition).
The notations used have their usual meanings. Cluster's heliocentric distance
is denoted by $d$ and the Galactocentric distance is denoted by $R_{GC}$.}
\centering
\label{par}
\begin{tabular}{cc}
\hline\hline Parameters & Values \\ \hline
$\alpha_{2000}$         & 10$^{\rm h}$ 17$^{\rm m}$ 36$^{\rm s}$.82  \\
$\delta_{2000}$         & $-46^\circ$ 24$\arcmin$ 44.9$\arcsec$    \\
$l$                     & $277\fdg23$                \\
$b$                     & $8\fdg64$              \\
$\rm [Fe/H]$            & $-1.59 $           \\
$E(B-V)  $              & 0.24 mag  \\
$d$                     & 4.9 kpc\\
$R_{GC}$                & 8.8 kpc\\
\hline
\end{tabular}
\end{table}
\begin{table}
\centering
\caption{Details of the WFI@{\bf 2.2}m archive data.
  The first epoch data were acquired on December 05, 1999
  and the second epoch data were taken between April 02-05, 2014.}
\label{log}
\begin{tabular}{cccc}
\hline
\hline
Filters      &  Exposure time & Seeing & Airmass \\
&(seconds)&(arcsec)& \\
\hline
\multicolumn{4}{c}{December 05, 1999 (First epoch)} \\
$B$&2$\times$240&1.1&$\sim$1.3     \\
$V$ &2$\times$240&0.9&$\sim$1.3     \\
$I$&2$\times$240&0.8&$\sim$1.2     \\
\multicolumn{4}{c}{April 02-05, 2014 (Second Epoch)} \\
$V$ &35$\times$40&1.1&$\sim$1.1     \\
\hline
\end{tabular}
\end{table}

\section{Data used and reduction procedures}
\label{OBS}

To determine the PMs of the stars in this work, we used archive
images\footnote{http://archive.eso.org/eso/eso\_archive\_main.html}
from observations made with the 2.2m ESO/MPI telescope at La Silla, Chile.
This telescope contains a mosaic camera called the Wide-Field Imager (WFI), consisting 
of 4$\times$2, i.e. 8 CCD chips.
Since each CCD has an array of 2048$\times$4096 pixels, WFI ultimately
produces images with a 34$\times$33 arcmin$^2$ field of view.

Table~\ref{log} presents the details of the observational log of the archive data.
The observational run of the first epoch
contain 2 images in $B, V$ and $I$ bands each with 240 sec exposure time
observed on December 05, 1999. In the second epoch, we have
35 images with 40 sec exposure time each
in $V$ filter observed between April 02$-$05, 2014.
Thus, the epoch gap between the data is $\sim$14.3 years.
As can be seen in the Table~\ref{log}, seeing
in the images used are 0.8--1.1 arcsec,
and airmass values lie between $\sim1.1$--1.3.

\subsection{The data reduction procedures}

To derive PMs from the WFI@2.2m mosaic CCD images, we used the
astrometric procedure developed by Anderson et al. (2006, hereafter A06).
The technique involves the usual initial steps of
de-biasing and flat-fielding.
One of the decisive factors in providing precise positions of the stars is constructing a good
Point Spread Function (PSF) for the WFI images.
Since the shape of the PSF changes across the mosaic CCD, we
capture this variability by using an array of 15 PSFs per CCD chip
(3 across and 5 high), as explained in A06.
To furnish  positions and fluxes of the objects in an image,
an array of $empirical$ PSFs are constructed.
These PSFs are saved in a look-up table on a very fine grid of a quarter pixel size.
Each PSF goes out to a radius of 25 pixels and each pixel is split in 4 equal parts,
thus giving (201, 201) grid points for a PSF.
The center of the PSF is located at the central gridpoint (101, 101).
A06 presented the automated code we are now using to iteratively determine the precise
positions and instrumental magnitudes for the brightest down to faintest stars for $B, V, I$ bands.

As documented in A06, WFI@{\bf 2.2}m is affected by significant
geometric distortion in the focal plane, which
leads the pixel scale across the field-of-view to change effectively.
The corrections to account for the geometric distortion were derived using
dithered observations of Baade's window which lies in the Galactic bulge (see, A06).
The corrections have been noted in a look-up table comprising
9$\times$17 elements, for each chip.
For any particular location, a bi-linear interpolation between the
four closest grid points from the look-up table to the target point delivers the
distortion correction. Still, the distortion may vary over time 
for the WFI, 
and is typically larger near the edges of 
the image.
These factors lead to uncertainty in distortion corrections.
To tackle this uncertainly, we followed
the local transformation approach
as described in Section 7 of the article A06.
According to the local transformation approach, transformations from one frame to another
are obtained locally, i.e. with respect to some stars in our own images.
Because cluster stars exhibit lesser amount of internal dispersion than the
field stars, cluster stars based on their location in CMD and motion 
are chosen as reference. Initially, we choose stars lying on the 
main sequence, sub giant and red giant branches by making blue and red envelope for 
the sequences in the CMD.
In subsequent attempts, the selection is done 
using PMs. The process is iterated
multiple times to provide the best possible results.
After positions of stars are determined in all frames, we use
six-parameter linear transformations
to transform the positions from one frame to another.
This approach resembles the classical ``plate-pair'' method 
(e.g., Sanders 1971a, Tian et al. 1998),
but it is more generalized and
can be used to all possible combinations of the first and second epoch frames.
The relative PM of a target star will be the
average of all displacements for inter epoch transformations.
Since PMs do not contribute to the intra-epoch displacements, they are
used to calculate errors in PM measurements.

\subsubsection{Calibrating the photometry}

Instrumental $B, V$ and $I$ magnitudes were transformed into standard Johnson--Cousin system using
secondary standard stars 
provided by
P. Stetson\footnote{http://www3.cadc-ccda.hia-iha.nrc-cnrc.gc.ca/community/STETSON/standards/}.
The standard stars used for calibration have a brightness range of  12.6$\le V \le$19.8
and color ranges of $0.4 \le (B-V) \le 1.5$ and $0.4 \le (V-I) \le 1.9$.
A total of 160 stars for $BV$ magnitudes and 165 stars 
for $I$ magnitudes were used in the calibration process.

We used the transformation equations written below to derive the photometric zero-points and color terms:\\

\begin{center}
$ B_{\rm std} = B_{\rm ins} + C_b\times(B_{\rm ins} - V_{\rm ins}) + Z_b $

$ V_{\rm std} = V_{\rm ins} + C_v\times(B_{\rm ins} - V_{\rm ins}) + Z_v $

$ I_{\rm std} = I_{\rm ins} + C_i\times(V_{\rm ins} - I_{\rm ins}) + Z_i $,\\
\end{center}
where instrumental magnitudes and secondary standard magnitudes have been denoted by
subscripts ``ins'' and ``std'' respectively.
$C_b, C_v$ and $C_i$ denote
the color terms, while $Z_b, Z_v$ and $Z_i$ are the global zero-points.
As a result of the calibration, the values of the color-terms are 0.39, $-$0.07 and 0.12,
whereas the zero-points are 24.79, 24.18, 23.27 for  $B$, $V$ and $I$ filters respectively.
These values of color terms and zero-point agree with the values posted
on the WFI@{\bf 2.2}m
webpage\footnote{http://www.ls.eso.org/lasilla/sciops/2p2/E2p2M/WFI/zeropoints/}.

The photometric standard deviations for individual photometric bands
were calculated by reducing multiple observations to a common reference frame.
Figure~\ref{errormag} presents
the rms error in the  magnitudes for $B, V$ and $I$ magnitudes
as a function of visual magnitude.
The values of average rms are less than $\sim$0.01 mag for stars brighter than 19 mag for
$B$ and $I$ filters and better than $\sim$0.01 mag for $V<20$ mag.

\subsubsection{Calibrating the positions}

As a part of the astrometric studies of NGC 3201,
we present the equatorial coordinates of stars
in International Celestial Reference System (ICRS).
We used the geometric distortion correction
from the look-up table given in the A06 to correct
the pixel coordinates $X, Y$ of each star in each frame
and averaged by means of a six parameter linear transformation into a common reference frame.
The online digitized sky ESO catalogue in the SKYCAT software is then used 
to transform the averaged
$X, Y$ positions to right ascension (RA) and declination (Dec) in J2000.0 equinox
using IRAF\footnote{IRAF is distributed by the
National Optical Astronomical Observatory which is operated by the
Association of Universities for Research in Astronomy, under contact with
the National Science Foundation} tasks $CCMAP$ and $CCTRAN$.
The transformations have rms values of about $\sim20$ mas.
The relatively high accuracy of our distortion corrections as well as
the reasonable stability of
the intra-chip positions makes it possible to apply a single plate model which includes
linear and quadratic terms and a small but significant cubic term in each coordinate.
Also, this solution removes the effects caused by differential refraction.

\begin{figure*}
\centering
\includegraphics[width=\textwidth]{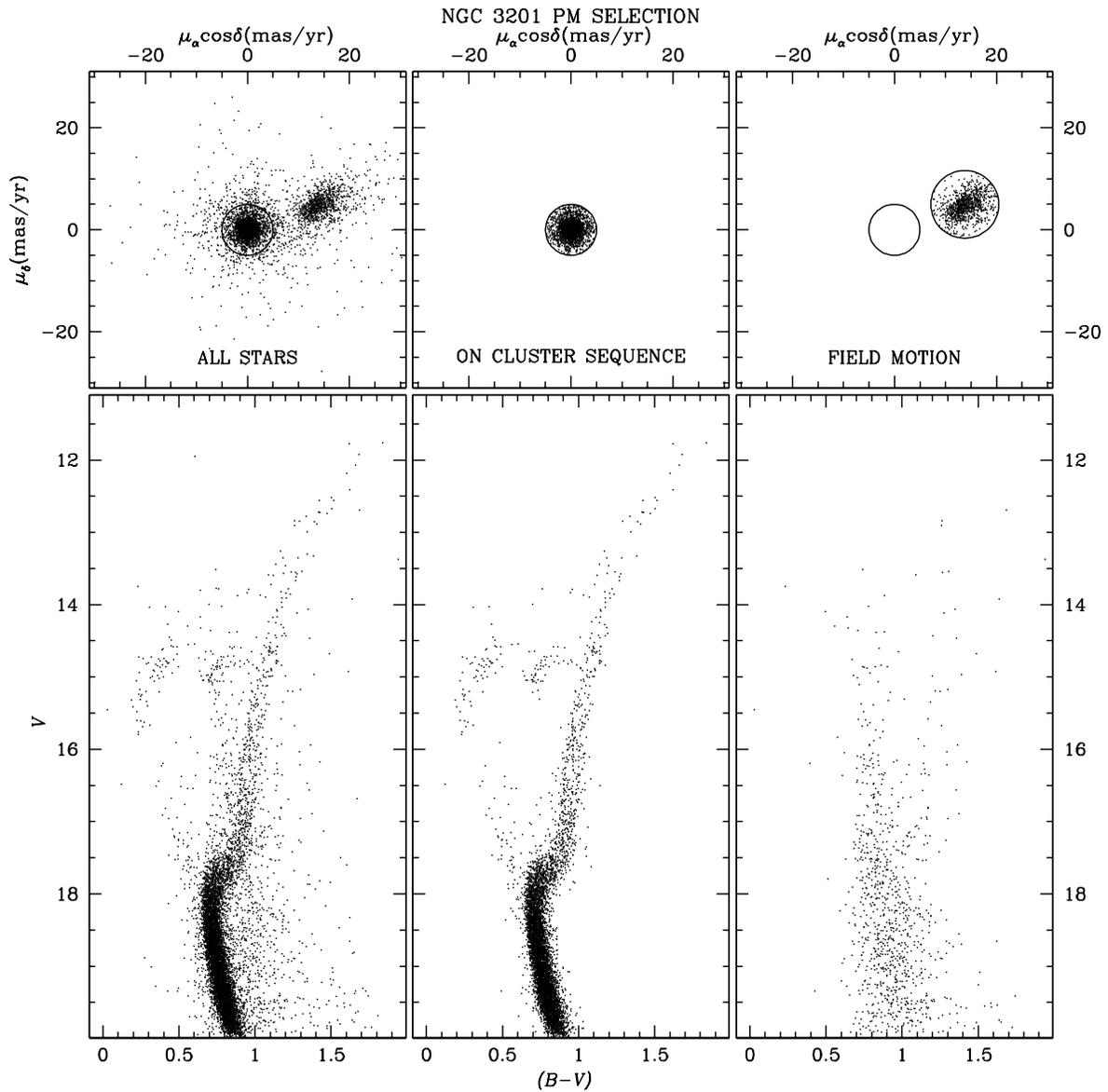}
\caption{
        {\em (Top panels)} Proper motion vector-point diagrams (VPDs).
        Zero point in VPD is the average motion of the assumed cluster stars.
        {\em (Bottom panels)}
$V$ vs. $(B-V)$ CMDs.
        {\em (Left)} The whole sample;
        {\em (center)} stars in VPD within $\sim$5 mas~yr$^{-1}$
         around the cluster mean motion.
        {\em (Right)} Probable field stars in the region
        of NGC~3201.
        For all the plots only stars 
	having PM error
        better than 2 mas~yr$^{-1}$
	in each
        coordinate have been considered.  }
\label{cmd_inst}
\end{figure*}
%
\begin{figure*}
\centering
\includegraphics[width=\textwidth]{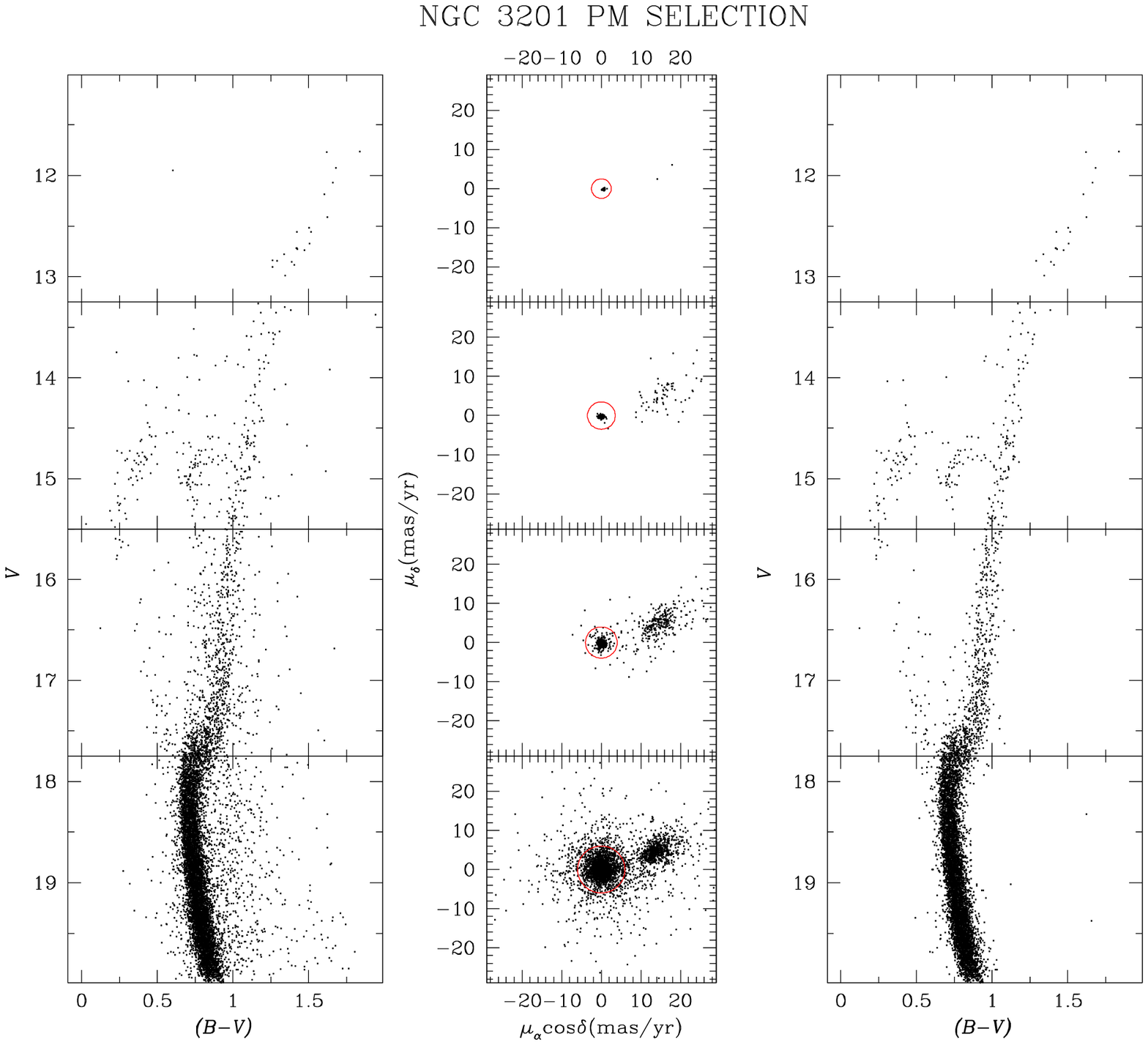}
\caption{
        {\em (Left:)} Magnitude-binned CMD of the stars while PM errors increase from 1.2 mas~yr$^{-1}$ for the brightest bin to
        2.5 mas~yr$^{-1}$ for the faintest bin.
        {\em (Middle:)} VPDs for the same stars lying in the
        corresponding magnitude bins. A circle in each VPD
        shows the adopted membership criterion.
        The radii of the circles increase from 2.5 mas~yr$^{-1}$ to 6.0 mas~yr$^{-1}$ from bright to fainter bins.
        {\em (Right:)}  CMD for stars expected
        to be cluster members.
        }
\label{cmd_II}
\end{figure*}

\subsection{Proper motion determinations}
\label{PM}

We used 6 images from the first epoch and 35 images from the
second epoch to determine PMs for NGC 3201 stars. 
Having a large number of images minimizes the value of the 
standard error in the PMs.

We started with a selection of photometric clusters members,
i.e. members selected on the basis of
their position in $V$ vs $(B-V)$ CMD.
We selected stars lying near the cluster sequences in the CMD, with
brightness in the range
of 14$\le V \le$19.
These are used as a local reference to transform the coordinates
of the stars between the epochs.
Adopting only those stars lying on the cluster sequences in the CMD and 
having PM errors $<$1.0 mas~yr$^{-1}$ ensure that
the PMs are determined with respect to the systematic motion of the cluster.
To minimize the influence of uncorrected geometric distortion residuals,
a local transformation based on the closest 25 reference stars
on the same CCD chip was used.
We did not find any systematics larger than random errors
close to the corners or edges of the CCD chips.

The routine described in A06 has an iterative nature and we
iterated it to remove some stars from the initial photometric
member list. 
Stars were eliminated if they had PMs inconsistent with cluster
membership, in spite of
their colors placing them near the
cluster sequence in the CMD. PMs and 
their rms errors are plotted as a function 
of visual magnitude in Figure~\ref{errorpm}.
The median value of PM error
is $\sim$0.8 mas~yr$^{-1}$ for stars brighter than $V\sim$18 mag 
which increases up to $\sim$1.5 mas~yr$^{-1}$ for stars
in the magnitude range of $18<V<20$ mag.

\subsubsection{Cluster CMD decontamination}

One of the main reasons to carry out PM analysis for star clusters
is to isolate the cluster sample from the field stars
and to produce a CMD with only the most probable cluster members.
Figure~\ref{cmd_inst} clearly demonstrates the strength of PM analysis in
separating the field stars using vector-point diagrams (VPDs) in the top panels,
in combination with $V$ versus $(B-V)$ CMDs in the bottom panels.
In the left panels of the figure, the entire sample of stars is shown,
while the middle panels and right panels represent likely cluster members
and field stars respectively.
In the top middle panel, VPD has a circle of radius $\sim$5 mas~yr$^{-1}$
around the cluster centroid.
This motion circle is our provisional criterion for assigning membership 
to the stars,
before membership probabilities are determined. 
The radius of the circle is chosen as a compromise between losing
cluster members with poorly measured PMs 
and including some field stars that have their PMs 
consistent with the cluster's mean PM.
The shape of the cluster members' PM distribution in the VPD is round,
which suggests that our PM measurements are not affected by any systematics.
It is pretty obvious from the figure that having a large epoch gap for CCD data
has produced a CMD almost free from field stars.

Figure~\ref{cmd_II} shows the $(B-V), V$ CMD which is binned along the magnitude axis.
To identify provisional cluster members, different selection criteria 
were used in different magnitude bins.
The criterion was tighter for bright stars as they have more reliable 
measurements, but is less stringent for fainter stars. As can be seen
in Figure~\ref{cmd_II}, fainter stars have poorer PM determinations compared 
to brighter stars: PM uncertainty increases
from 1.2 mas yr$^{-1}$ to 2.5 mas yr$^{-1}$
from the brightest bin to the faintest one. 
We therefore adopt a PM selection radius that increases from
2.5 mas~yr$^{-1}$ in the brightest magnitude bin to 6.0 mas~yr$^{-1}$ 
in the faintest one. However, we still
have a good enough decontamination of field stars even 
in the fainter magnitudes. 

\section{Membership probabilities}
\label{MP}

In Figure~\ref{cmd_inst}, two different groups of stars are distinguishable
based on their motion, although a larger fraction of the stars are
inside the circle for provisional cluster membership.
The next step is to determine membership probabilities of stars which will yield a
quantitative significant number for a particular star belonging to the cluster.
The credit to set up a mathematical model to use PMs to determine membership probabilities
goes to Vasilevskis et al. (1958).
The maximum likelihood principle to compute membership 
probabilities was introduced by Sanders (1971b). Over the years,
methods to calculate membership probabilities have been refined
(e.g., Stetson 1980;  Zhao \& He 1990; Zhao \& Shao 1994).
In this study, we use the 
method given by Balaguer-N\'{u}\~{n}ez et al. (1998).
This method has been previously used for both globular clusters
(Bellini et al. 2009, Sariya et al. 2012, 2015) and open clusters (Yadav et al. 2013).
According to this method, two frequency distribution functions are constructed for
a particular i$^{th}$ star. Frequency distributions of cluster stars
($\phi_c^{\nu}$) and field stars ($\phi_f^{\nu}$) are presented by the equations given below:\\

\begin{center}
   $\phi_c^{\nu} =\frac{1}{2\pi\sqrt{{(\sigma_c^2 + \epsilon_{xi}^2 )} {(\sigma_c^2 + \epsilon_{yi}^2 )}}}$

$\times$ exp$\{{-\frac{1}{2}[\frac{(\mu_{xi} - \mu_{xc})^2}{\sigma_c^2 + \epsilon_{xi}^2 } + \frac{(\mu_{yi} - \mu_{yc})^2}{\sigma_c^2 + \epsilon_{yi}^2}] }\}$ \\
\end{center}
\begin{center}
and\\
\end{center}
\begin{center}
$\phi_f^{\nu} =\frac{1}{2\pi\sqrt{(1-\gamma^2)}\sqrt{{(\sigma_{xf}^2 + \epsilon_{xi}^2 )} {(\sigma_{yf}^2 + \epsilon_{yi}^2 )}}}$

$\times$ exp$\{{-\frac{1}{2(1-\gamma^2)}[\frac{(\mu_{xi} - \mu_{xf})^2}{\sigma_{xf}^2 + \epsilon_{xi}^2}}
-\frac{2\gamma(\mu_{xi} - \mu_{xf})(\mu_{yi} - \mu_{yf})} {\sqrt{(\sigma_{xf}^2 + \epsilon_{xi}^2 ) (\sigma_{yf}^2 + \epsilon_{yi}^2 )}} + \frac{(\mu_{yi} - \mu_{yf})^2}{\sigma_{yf}^2 + \epsilon_{yi}^2}]\}$\\
\end{center}

where ($\mu_{xi}$, $\mu_{yi}$) are the PMs of $i^{th}$ star, while
($\epsilon_{xi}$, $\epsilon_{yi}$) are the proper motion errors.
($\mu_{xc}$, $\mu_{yc}$) represent the cluster's PM
center and ($\mu_{xf}$, $\mu_{yf}$) are the field PM center.
For the cluster members, the intrinsic proper motion dispersion is denoted by
$\sigma_c$, whereas $\sigma_{xf}$ and $\sigma_{yf}$ exhibit the field
intrinsic proper motion dispersions. The correlation coefficient $\gamma$ is calculated as:\\

\begin{center}
$\gamma = \frac{(\mu_{xi} - \mu_{xf})(\mu_{yi} - \mu_{yf})}{\sigma_{xf}\sigma_{yf}}$.
\end{center}
\begin{figure}
\vspace{-3.0cm}
\centering
\includegraphics[width=8.5cm]{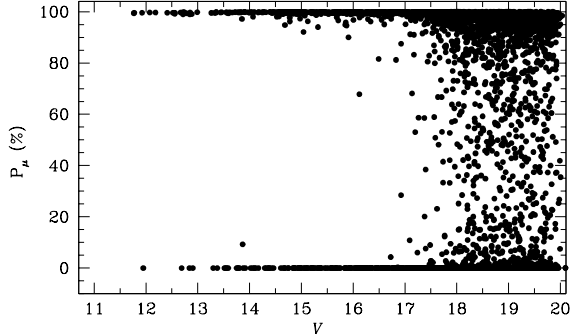}
\caption{Membership probability P$_{\mu}(\%)$ of the stars in the direction of NGC~3201 plotted as a function  of the  $V$ magnitude.
For stars having magnitudes $V \sim 18$ mag and fainter,
the average P$_{\mu}$ are decreasing for cluster members, while
they are increasing for field stars.} 
\label{VvsMP}
\end{figure}
\begin{figure}
\centering
\includegraphics[width=8.5cm]{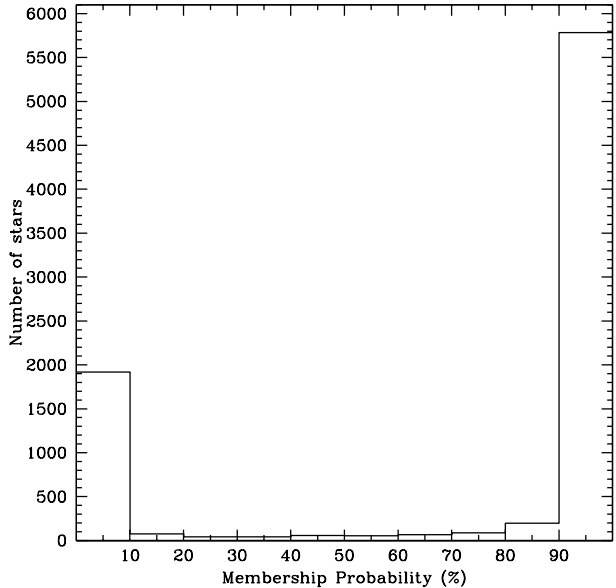}
\caption{Histogram of the membership probabilities derived in this study.
}
\label{hist}
\end{figure}
\begin{figure}
\centering
\includegraphics[width=8.5cm]{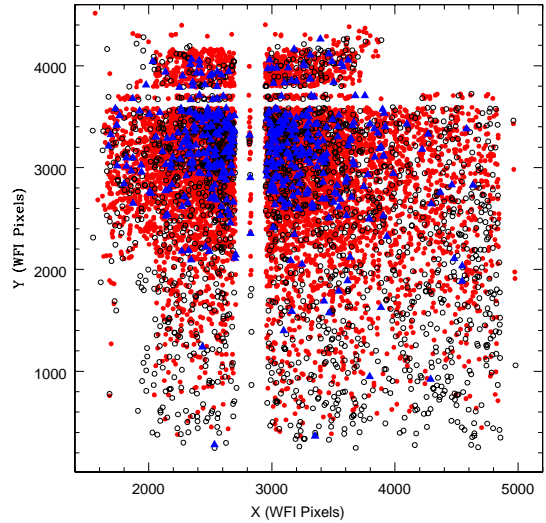}
\caption{Spatial distribution of stars presented in this study. 
Red filled circles are the stars with P$_{\mu} > 80\%$, blue triangles show the stars having $10\%<P_{\mu} \leq 80\%$,
and black open circles show the stars with P$_{\mu} \leq 10\%$. Stars having proper motion error better than 
2 mas yr$^{-1}$ are plotted here.
The gaps which form a ``cross'' like pattern are due to the gaps in the mosaic CCD system.}
\label{spatialmp}
\end{figure}
\begin{figure}
\centering
\includegraphics[width=8.5cm]{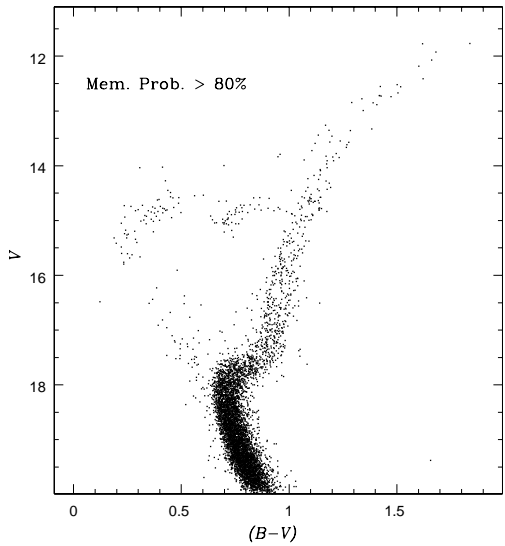}
\caption{$V$ vs. $(B-V)$ CMD for the stars with 
membership probabilities $>$ 80\%} 
\label{mp_cmd}
\end{figure}
\begin{figure}
\centering
\includegraphics[width=8.5cm]{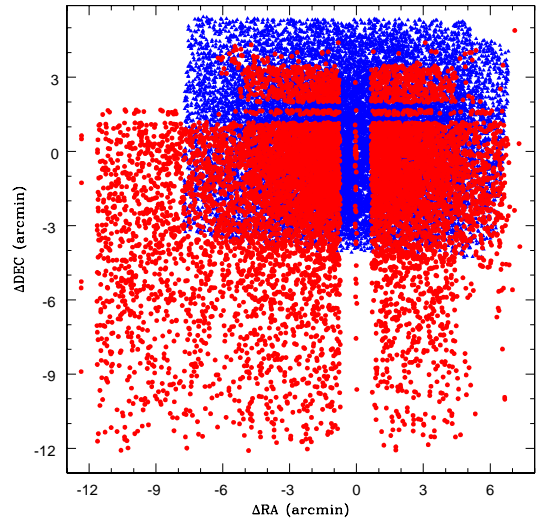}
\caption{Spatial distribution of stars in the present and Zl12 catalogues.
Axes shown here are the differences in coordinates of NGC 3201 stars and the cluster 
center in arcmin. Red filled circles show the present catalogue while blue triangles 
represent the catalogue of Zl12.}
\label{spatialboth}
\end{figure}
\begin{figure}
\centering
\includegraphics[width=8.5cm]{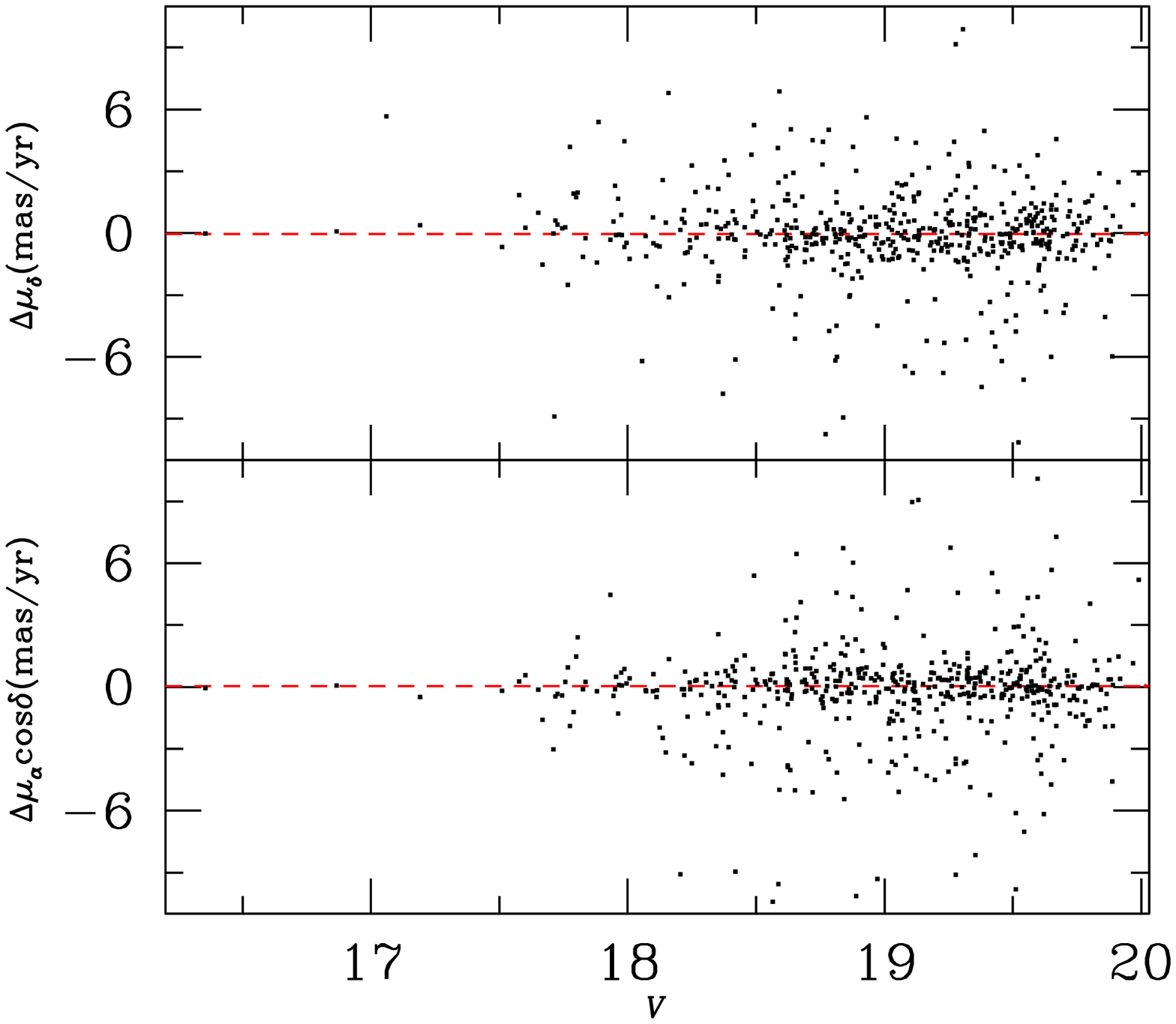}
\caption{Difference between the Zl12 proper motions and those derived in 
the present study, plotted against visual magnitude.
3$\sigma$-clipped median PM differences are shown by
the horizontal dashed lines.}
\label{kamil}
\end{figure}

The spatial distribution of the stars was not considered 
in calculating membership.
In computing $\phi_c^{\nu}$ and $\phi_f^{\nu}$, we used those stars which have 
PM errors better than $\sim$2 mas~yr$^{-1}$.  In VPD, the center of the cluster stars
is found to be
($\mu_{xc}$, $\mu_{yc}$)=(0, 0) mas~yr$^{-1}$. 
The  intrinsic PM dispersion for the cluster stars ($\sigma_c$)
could not be ascertained reliably using our PM data. 
Pryor \& Meylan (1993) list the value of radial velocity dispersion for NGC~3201
as 5.2 km~sec$^{-1}$. Considering the value of the distance of NGC 3201 as
4.9 kpc (Harris 1996, 2010 edition),
the internal PM dispersion becomes $\sim$0.22 mas~yr$^{-1}$.
Hence, we used $\sigma_c$= 0.22 mas~yr$^{-1}$.
For field stars, we have
($\mu_{xf}$, $\mu_{yf}$) = (13.8, 4.8) mas yr$^{-1}$ and
($\sigma_{xf}$, $\sigma_{yf}$) = (2.5, 2.1) mas yr$^{-1}$.  \\

If $n_{c}$ and $n_{f}$ are the normalized number of cluster and field stars respectively
(i.e., $n_c + n_f = 1$), the total distribution function can be calculated as:\\

\begin{center}
$\phi = (n_{c}~\times~\phi_c^{\nu}) + (n_f~\times~\phi_f^{\nu})$,  \\
\end{center}

As a result, the membership probability for the $i^{th}$ star is given by:\\
\begin{center}
$P_{\mu}(i) = \frac{\phi_{c}(i)}{\phi(i)}$. \\
\end{center}

In Figure~\ref{VvsMP}, 
membership probabilities are shown as a function of visual magnitude.
The figure shows clear separation of cluster
and field stars as sharp distributions of stars 
around membership values $P_{\mu}\sim$100$\%$ and  $P_{\mu}\sim$0$\%$.
However, due to increasing errors at fainter regime,
one can notice a number of stars with 
intermediate values of  $P_{\mu}$, for magnitudes fainter than $V$=18 mag.
The histogram of membership probabilities for 8322 stars is shown in Figure~\ref{hist}.
The presence of higher peaks for the
leftmost and rightmost bins suggest that 
the method used for the  membership
determination is effective for NGC~3201. 
We find 5981 stars that have 
membership probabilities larger than 80$\%$.
The spatial distribution of the stars is shown in Figure~\ref{spatialmp}.
To distinguish between the cluster members and field stars,
we have used different symbols for stars having  P$_{\mu} > 80\%$,
$10\%<P_{\mu} \leq 80\%$ and P$_{\mu} \leq 10\%$.
The cut off values of P$_{\mu}$ are based on the histogram shown in Figure~\ref{hist}.

Figure~\ref{mp_cmd} presents the CMD of stars 
having P$_{\mu} > 80\%$.
In this CMD, cluster sequences for stars brighter than $V \sim 20$ mag 
can be seen. Also, this CMD shows stars of various
evolutionary stages like sub-giants, red giants, horizontal branch 
stars and blue stragglers. 
All the cluster sequences in this CMD
look cleaner with minimal field contamination.
As mentioned earlier, NGC 3201 shows 
differential reddening, due to which the cluster sequences
are broadened in CMD. As discussed by Kravtsov et al. (2009), 
the red giant branch of the cluster, in particular, shows significant
broadening. In addition to this, we find gaps in the red giant branch at $V\sim$13 
and $V\sim$15, the former one being more significant. 
These gaps have been reported by
Lee (1977) in his photometric study of NGC 3201.
The horizontal branch of NGC 3201 is well developed and extended, with
a relatively unclear instability strip.

\subsection{Comparison with Zl12}
The PMs presented in this study were compared with the catalogue given by Zl12.
For comparison, we plotted the spatial distributions of our catalogue with Zl12
in Figure~\ref{spatialboth}. In this figure, our catalogue stars are shown with red 
filled circles, while Zl12 stars are shown with blue triangles. It is clearly seen 
in the figure that the present investigation 
extends the PM studies of NGC 3201 to a wider region. 
In the additional observed area in our catalogue, we have PM 
information from about 2000 stars.
For the common stars, the differences in both PM components between the two catalogues is 
shown in Figure~\ref{kamil}.
Values of the 3$\sigma$-clipped median of the proper motion differences are
$0.04 (\sigma = 0.54)$ mas yr$^{-1}$ and $-0.05 (\sigma = 0.62)$ mas yr$^{-1}$.
Our PMs exhibit consistency with the Zl12 data for $V<20$ mag.

\section{Membership of variables and $X$-ray sources}
\label{app}
\begin{table*}
\centering
\caption{Membership probability of the known variables (ID$_V$) in the direction of NGC 3201 and
$X$-ray sources from Webb et al. (2006) (ID$_X$). Star number in our catalogue is denoted by ID.}
\begin{tabular}{ccc|ccc}
\hline
& Variables  &     &   &  $X$-ray sources   &         \\
\hline
ID$_V$  &  P$_{\mu}$   & ID        &  ID$_X$  &  P$_{\mu}$    & ID \\
&(\%) & & &(\%) &\\
\hline
V71  &   0  & 3373   & 22  & 100  &  8134  \\
V92  & 100  & 5147   & 25  &   2  &  6766  \\
V98  & 100  & 4921   & 26  & 100  &  6374  \\
V137 &   6  & 7804   & 30  &   0  &  1963  \\
     &      &        & 33  &   0  &   624  \\
     &      &        & 45  &  99  &  8222  \\
     &      &        & 46  &   1  &  3222  \\
     &      &        & 48  &  77  &   702  \\
     &      &        & 58  & 100  &  7888  \\
     &      &        & 59  &  98  &  1229  \\
\hline
\label{var01}
\end{tabular}
\end{table*}
\begin{figure}
\centering
\includegraphics[width=8.5cm]{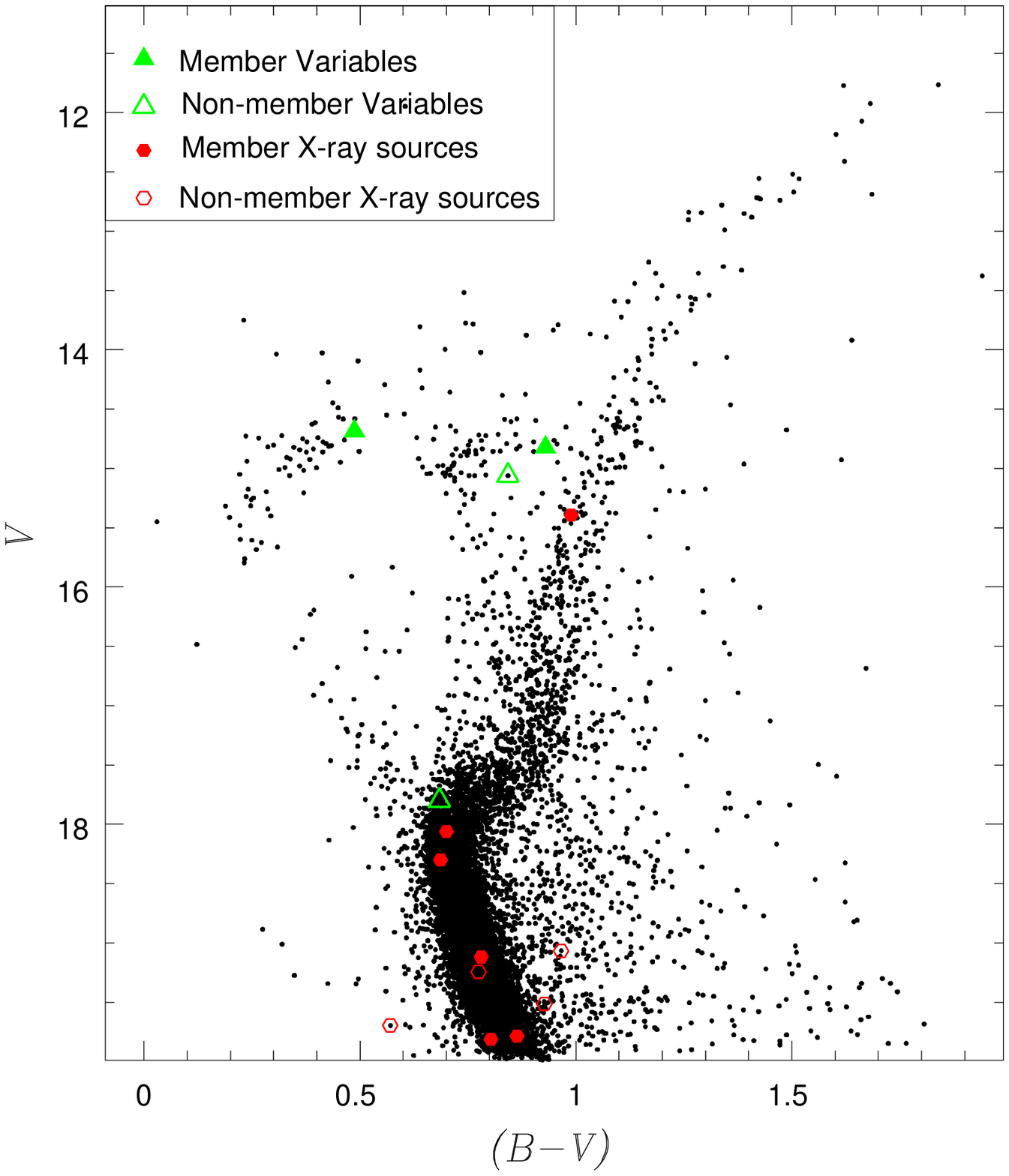}
\caption{The variable stars and $X$-ray sources found in our membership catalogue are shown in this CMD with
different symbols as indicated in the inset.
Filled circles and filled triangles show the most probable cluster members, while the open circles
and open triangles show the kinematic non-members.
This CMD represents all the stars for which membership probabilities have been calculated.}
\label{cmd_vari}
\end{figure}

We used the membership catalogue to ascertain the membership status of the reported variable stars and $X$-ray sources
in the region of NGC 3201. The details of the comparison are listed in Table~\ref{var01}.
The variable stars of NGC 3201 have been compiled on Clement's webpage of the catalogue of variable stars in
globular clusters\footnote{http://www.astro.utoronto.ca/$\sim$cclement/cat/C1015m461}.
Recently, some new variable were detected by Kaluzny et al. (2016).
We found four variable stars in common with our catalogue.
In Table~\ref{var01}, V137 comes from the variables listed in Kaluzny et al. (2016) and the coordinates of the other three variables
have been taken from Clement's webpage.
As can be seen in Table~\ref{var01}, variables V92 and V98 have a 100\% probability of
being cluster stars. The two other variable stars, V71 and V137 have $P_{\mu}$$\sim$0\%,
which implies that they are not cluster members.
The $X$-ray sources in the NGC 3201 field-of-view have been identified by
Webb et al. (2006). Out of 10 $X$-ray sources found, five sources
(ID$_X$= 22, 26, 45, 58, 59) have membership probability close to 100\%, which makes them most probable cluster members.
Another $X$-ray source, ID$_X$= 48 should also be a probable cluster member based on its membership probability.
ID$_X$= 25, 30, 33, 46 have  $P_{\mu}$$\sim$0\%, 
which means they do not belong to NGC 3201.
Figure~\ref{cmd_vari} shows the CMD of all the stars listed in our membership 
catalogue with the variables and X-ray sources shown with different symbols.
\section{The catalogue}
\label{catl}

The electronic catalogue from the present study contains
relative PMs, PM errors, membership probabilities,
photometric magnitudes in $B, V, I$ bands with rms errors for 8322 stars in the direction of
NGC~3201. Column (1) of the catalogue gives the running number; Columns. (2) and (3) present the
J2000.0 equatorial coordinates, while Columns. (4) and (5) show positions 
of the stars ($X$ and $Y$) in CCD pixels.
Relative PMs and PM errors are listed in Columns. (6) to (9).
Columns (10) to
(15) provide photometric $B, V$ and $I$ magnitudes and their
rms errors.
Final Column, column number (16) represent the membership probability $P_{\mu}$.
Some initial lines from the electronic catalogue are shown in Table \ref{cata}.

\begin{sidewaystable*}
\centering
\caption{~The membership catalogue for NGC~3201 (initial few lines).}
\vspace{0.4cm}
\tiny
\begin{tabular}{cccccccccccccccc}
\hline\hline
ID & $\alpha_{2000}$ & $\delta_{2000}$ &X&Y & $\mu_{\alpha}cos(\delta)$ &$ \sigma_{\mu_{\alpha}cos(\delta)}$ &$\mu_{\delta}$ & $\sigma_{\mu_{\delta}}$&B&$\sigma_B$&V&$\sigma_V$&I&$\sigma_I$&$P_{\mu}$\\
(1)&(2)&(3)&(4)&(5)&(6)&(7)&(8)&(9)&(10)&(11)&(12)&(13)&(14)&(15)&(16) \\
&[h:m:s]&[d:m:s]&[pixel]&[pixel]&[mas/yr]&[mas/yr]&[mas/yr]&[mas/yr]&[mag]&[mag]&[mag]&[mag]&[mag]&[mag&[$\%$]\\
\hline
1 & 10:17:43.87 & -46:36:49.1 & 2535.4886 & 247.4757 & 10.9367 & 0.7142 & 2.9779  & 0.3006 & 17.4382 & 0.0044 & 16.6389 & 0.0045 & 15.7423 &  0.0052 &   0\\
 2 & 10:17:17.45 & -46:36:50.5 & 3681.3510 & 247.7929 & -0.6062 & 1.0181 & 1.4117  & 0.1478 & 18.8624 & 0.0022 & 18.1265 & 0.0017 & 17.2219 &  0.0008 &   0\\
 3 & 10:16:54.53 & -46:36:49.9 & 4675.0902 & 254.5138 & 14.3016 & 0.8902 & 0.8736  & 1.3054 & 17.4766 & 0.0028 & 16.5418 & 0.0061 & 15.4859 &  0.0041 &   0\\
 4 & 10:17:27.99 & -46:36:46.7 & 3224.0258 & 261.3223 & 13.9711 & 0.8437 & 6.6451  & 1.6193 & 19.7695 & 0.0055 & 18.7313 & 0.0111 & 17.6100 &  0.0029 &   0\\
 5 & 10:17:06.37 & -46:36:44.3 & 4161.2320 & 275.8470 & 14.7467 & 0.6278 & 9.3091  & 0.4916 & 17.9729 & 0.0044 & 16.8013 & 0.0039 & 15.4686 &  0.0024 &   0\\
 6 & 10:16:55.80 & -46:36:44.2 & 4619.5001 & 278.2980 & 10.9301 & 0.8520 & 4.0641  & 1.6559 & 20.1599 & 0.0077 & 19.1355 & 0.0094 & 17.9470 &  0.0163 &   0\\
 7 & 10:17:43.89 & -46:36:41.3 & 2534.5683 & 279.9379 & -0.9899 & 1.1377 & 2.5727  & 0.9101 & 19.7797 & 0.0104 & 18.9841 & 0.0012 & 18.0374 &  0.0085 &  45\\
 8 & 10:17:41.74 & -46:36:37.4 & 2627.5263 & 297.0895 & 13.4994 & 1.0762 & 5.7681  & 1.3270 & 19.7509 & 0.0060 & 18.6114 & 0.0076 & 17.3379 &  0.0075 &   0\\
 9 & 10:17:49.71 & -46:36:36.2 & 2281.8995 & 299.8237 & 13.6555 & 1.1161 & 1.7871  & 1.8352 & 19.1304 & 0.0055 & 18.0739 & 0.0017 & 16.9377 &  0.0057 &   0\\
10 & 10:17:22.46 & -46:36:31.1 & 3463.3688 & 328.1909 & 13.0958 & 0.5066 & 5.6286  & 0.2441 & 18.3758 & 0.0104 & 17.3988 & 0.0039 & 16.3546 &  0.0008 &   0\\
11 & 10:17:28.71 & -46:36:29.1 & 3192.3215 & 335.1330 & 10.8703 & 1.1676 & 6.7231  & 1.3137 & 18.7126 & 0.0077 & 17.8893 & 0.0138 & 16.9378 &  0.0057 &   0\\
\hline
\label{cata}
\end{tabular}
\end{sidewaystable*}
\section{Conclusions}
\label{con}

We provide proper motions and membership probabilities for 8322 stars
in the field of globular cluster NGC~3201.
Our PMs successfully separate field stars from cluster members, and we
present a CMD which shows cluster sequences almost free from field stars.
An electronic catalogue with equatorial coordinates, PMs, membership probability
and $BVI$ photometry is being made available to the astronomical community.
The membership catalogue has been used to define the membership status of
variable stars and $X$-ray sources earlier reported in the cluster's direction.
This work also demonstrates that CCD data with only few years
epoch gap can provide proper motions accurate enough for kinematic
separation of cluster members.

\acknowledgements
We are indebted to an anonymous referee for a careful reading of the
manuscript and giving very useful suggestions which greatly improved
this article including its language and presentation.
Devesh P. Sariya and Ing-Guey Jiang acknowledge the grant from
Ministry of Science and Technology
(MOST), Taiwan. The grant numbers are MOST 103-2112-M-007-020-MY3,
MOST 104-2811-M-007-024, and MOST 105-2811-M-007 -038.
This research used the facilities of the Canadian Astronomy Data Centre operated by the
National Research Council of Canada with the support of the Canadian Space Agency.
We have also used the catalogue of variable stars managed by Prof. Christine Clement
from University of Toronto, Canada.


\begin{thebibliography}{}

\bibitem[Anderson et al. (2006)]{anderson06} Anderson, J., Bedin, L. R., Piotto, G., Yadav, R. K. S., Bellini, A., 2006, A\&A, 454, 1029, [A06]

\bibitem[Alcaino (1976)]{alcaino76} Alcaino, G., 1976, A\&AS, 26, 251

\bibitem[1981]{alcaino81} Alcaino, G., Liller, W., 1981, AJ, 86, 1480

\bibitem[1989]{alcaino89} Alcaino, G., Liller, W., Alvarado, F., 1989, A\&A, 216, 68

\bibitem[2015]{anguiano2015} Anguiano, B., Zucker, D. B., Scholz, R.-D., Grebel, E. K. et al., 2015, MNRAS, 451, 1229

\bibitem[2016]{anguiano2016} Anguiano, B., De Silva, G. M., Freeman, K., Da Costa, G. S. et al., 2016, MNRAS, 457, 2078

\bibitem[2014]{arellano} Arellano Ferro, A., Ahumada, J. A., Calder\'on, J. H., Kains, N., 2014, Rev. Mex. Astron. Astrofis., 50, 307

\bibitem[1998]{balaguer} Balaguer-N\'{u}\~{n}ez, L., Tian, K. P., Zhao, J. L., 1998, A\&AS, 133, 387

\bibitem[2009]{bellini} Bellini, A., Piotto, G., Bedin, L. R., et al., 2009, A\&A, 493, 959

\bibitem[1993]{brewer} Brewer, J. P., Fahlman, G. G., Richer, H. B. et al., 1993, AJ, 105, 2158

\bibitem[2007] {Cdinescu} Casetti-Dinescu, D. I., Girard, T. M., Herrera, D., van Altena, W. F. et al., 2007, AJ, 134, 195

\bibitem[1984a]{cacciaria} Cacciari, C., 1984a, AJ, 89, 231

\bibitem[1984b]{cacciarib}Cacciari, C., 1984b, AJ, 89, 1082

\bibitem[2010]{chen} Chen, C. W., Chen, W. P., 2010, ApJ, 721, 1790

\bibitem[1988]{chun} Chun, M.-S., 1988, The Journal of Korean Astronomical Society, 21, 67

\bibitem[1994]{cote94}Cote, P., Welch, D. L., Fischer, P., Da Costa, G. S., Tamblyn, P., Seitzer, P., \& Irwin, M. J., 1994, ApJ Suppl. Ser., 90, 83

\bibitem[1995]{cote95} Cote, P., Welch, D. L., Fischer, P., Gebhardt, K., 1995, ApJ, 454, 788

\bibitem[2003]{covey} Covey, K. R., Wallerstein, G., Gonzalez, G., Vanture, A. D., Suntzeff, N. B., 2003, PASP, 115, 819

\bibitem[1997]{covino} Covino, S., Ortolani, S., 1997, A\&A, 318, 40

\bibitem[1986]{cudworth86} Cudworth, K. M., 1986, IAUS, 109, 201

\bibitem[1997]{cudworth97} Cudworth, K. M., 1997, ASP Conf. Series, 127, 91

\bibitem[1993]{DaCosta} Da Costa, G. S., Tamblyn, P., Seitzer, P. et al., 1993, ASP Conf. Ser., 50, 81

\bibitem[2006]{dambis} Dambis, A. K., 2006, Astronomical and Astrophysical Transactions, 25, 185

\bibitem[1998]{gonzalez} Gonzalez, G., Wallerstein, G., 1998, AJ, 116, 765

\bibitem[1996]{harriswe} Harris, W. E., 1996, AJ, 112, 1487

\bibitem[2016]{kaluzny} Kaluzny, J., Rozyczka, M., Thompson, I. B., Narloch, W., Mazur, B., Pych, W. et al. 2016, Acta Astronomica, 66, 31

\bibitem[2009] {Kravtsov09} Kravtsov, V., Alca{\'{\i}}no, G., Marconi, G., Alvarado, F., 2009, A\&A, 497, 371

\bibitem[2013] {Kravtsov10} Kravtsov, V., Alca{\'{\i}}no, G., Marconi, G. Alvarado, F., 2010, A\&A, 512, L6

\bibitem[2014]{kunder} Kunder, A., Bono, G., Piffl, T., Steinmetz, M. et al., 2014, A\&A, 572, A30

\bibitem[2003]{layden}Layden, A. C., Sarajedini, A., AJ, 2003, 125, 208

\bibitem[1977]{lee} Lee, S.-W., 1977, A\&AS, 28, 409

\bibitem[1999]{LeeJ} Lee, J-W., Carney B. W., 1999, AJ, 118, 1373

\bibitem[2003]{2003} Mazur, B., Krzeminski, W., and Thompson, I. B., 2003, MNRAS, 340, 1205

\bibitem[1967]{menzies} Menzies, J., 1967, Ph.D. thesis, Australian National University

\bibitem[2013]{munoz}Mu{\~n}oz, C., Geisler, D., Villanova, S., 2013, MNRAS, 433, 2006

\bibitem[2015] {Mucciarelli} Mucciarelli, A., Lapenna, E., Massari, D., Ferraro, F. R., and Lanzoni, B., 2015, ApJ, 801, 69

\bibitem[1984]{penny} Penny, A.J., 1984, in Observational Tests of the Stellar Evolution Theory, IAU Symp. 105, ed(s). A. Maeder and A. Renzini (Dordrecht:Reidel), 157

\bibitem[2002]{piersimoni} Piersimoni, A. M., Bono, G., Ripepi, V., 2002, AJ, 124, 1528

\bibitem[2004]{piotto} Piotto, G., Bedin, L. R., Cassisi, S., Anderson, J. et al., 2004, Mem. S. A. It. Suppl., 5, 71

\bibitem[1993]{pryor} Pryor, C., Meylan, G., 1993, ASP Conf. Series, 50, 357

\bibitem[1971]{sanders} Sanders, W. L., 1971a, A\&A, 15, 368

\bibitem[1971]{sanders} Sanders, W. L., 1971b, A\&A, 14, 226

\bibitem[2012]{Sariya12} Sariya, D. P., Yadav, R. K. S., Bellini, A., 2012, A\&A, 543, A87

\bibitem[2015]{Sariya15} Sariya, D. P., Yadav, R. K. S., 2015, A\&A, 584, A59

\bibitem[2013] {simmerer} Simmerer, J., Ivans, I. I., Filler, D. et al., 2013, ApJL, 764, L7

\bibitem[1980]{stetson} Stetson, P. B., 1980, AJ, 85, 387

\bibitem [1998] {Tian} Tian, K.-P., Zhao, J.-L., Shao, Z.-Y.,  Stetson, P. B., 1998, A\&AS, 131, 89

\bibitem[1958]{vasilevskis} Vasilevskis, S., Klemola, A., Preston, G., 1958, AJ, 63, 387

\bibitem[2001]{vonbraun} von Braun, K., Mateo, M., 2001, AJ, 121, 1522

\bibitem[2002]{vonbraun1} von Braun, K., Mateo, M. 2002, AJ, 123, 279

\bibitem[2006]{webb} Webb, N. A., Wheatley, P. J., Barret, D., 2006, A\&A, 445, 155

\bibitem[2008]{yadav08} Yadav, R. K. S., Bedin, L. R., Piotto, G. et al., 2008, A\&A, 484, 609

\bibitem[Yadav et al. (2013)]{Yadav13} Yadav, R. K. S., Sariya, D. P., Sagar, R., 2013, MNRAS, 430, 3350

\bibitem[Zhao et al. (1990)]{zhao90} Zhao, J. L. \& He, Y. P., 1990, A\&A, 237, 54

\bibitem[Zhao et al. (1994)] {Zhao94} Zhao, J. L. \& Shao, Z. Y., 1994, A\&A, 288, 89

\bibitem[Zloczewski et al. (2012)]{Zloczewski12} Zloczewski, K., Kaluzny, J., Rozyczka, M., Krzeminski, W., Mazur, B., 2012, Acta Astronomica, 62, 357
\end{thebibliography}
\end{document}